\documentclass[a4paper,11pt]{article}

\usepackage{amsmath,amssymb}
\usepackage{amsthm}
\usepackage{graphicx}
\usepackage{xcolor}
\usepackage[round]{natbib} %revised
\usepackage[colorlinks=true, linkcolor=blue!60!black, urlcolor=blue!60!black, citecolor=blue!60!black]{hyperref}
\usepackage[nameinlink]{cleveref}
\usepackage{tikz}
\usetikzlibrary{arrows.meta,calc,fit,positioning}

\usepackage[margin=1in]{geometry}
\usepackage[onehalfspacing]{setspace}
\usepackage{fontspec}
\usepackage{newtxmath}
\usepackage{newpxtext}

\theoremstyle{definition}
\newtheorem{theorem}{Theorem}
\newtheorem{corollary}{Corollary}
\newtheorem{definition}{Definition}
\newtheorem{example}{Example}
\newtheorem{lemma}{Lemma}
\newtheorem{proposition}{Proposition}

\title{The Limits of Rank-Dependent Priorities in School Choice\thanks{I am grateful to Fuhito Kojima for his guidance. I thank Kenzo Imamura, Peiyang Li, Shunya Noda, Tohya Sugano, Kenji Utagawa, Alexander Westkamp, Yuichi Yamamoto, and participants of Game Theory Workshop 2026, Asian Game Theory Conference 2026, and 32nd Decentralization Conference for discussions and comments. All errors are my own.}}
\author{Masato Eguchi\thanks{Graduate School of Economics, the University of Tokyo. Email: \href{mailto:eguchi-masato967@g.ecc.u-tokyo.ac.jp}{\text{eguchi-masato967@g.ecc.u-tokyo.ac.jp}}}}
\date{\today}

\begin{document}
\maketitle
\begin{abstract}
    Many school choice mechanisms give students higher priority at schools they rank highly and aim to reward students' eagerness. This paper studies whether a centralized mechanism can incorporate such rank-dependent priorities while preserving truthful reporting. I consider a mechanism that jointly specifies how reported preferences modify schools' original priorities and how assignments are determined based on the modified priorities. I require the modification rule to satisfy rank monotonicity and the resulting matching to be stable with respect to the modified priorities. These two requirements jointly capture the policy objective of rewarding eagerness. My main result shows that any mechanism satisfying these requirements as well as strategy-proofness is outcome equivalent to the student-proposing Deferred Acceptance mechanism using the original priorities. Thus, if eagerness affects the outcome while the rank-dependent policy goal is maintained, the mechanism must be manipulable.
\end{abstract}

\section{Introduction}\label{sec:introduction}
The design of school choice mechanisms has long been studied and played an important role in admissions reforms in cities such as New York, Boston, and Chicago. In particular, the student-proposing Deferred Acceptance mechanism ($DA^S$) is celebrated for being both stable and strategy-proof and has been widely adopted in practice. These properties hold in the canonical model of school choice established by \citet{abdulkadiroglu2003}, in which schools' priorities reflect criteria such as test scores, residential proximity, or sibling status.
A central feature of this model is that priorities are determined independently of students' submitted preferences. Students' reports therefore affect the assignments without changing their relative priority at each school. \par
In practice, however, some assignment systems make each school's priority depend on students' submitted rankings, reflecting a policy objective of giving greater priority to students who are more eager to attend the school. The Boston mechanism favors students who rank a school more highly, while the Taiwan assignment mechanism deducts points from exam scores according to a school's position on the reported preference list \citep{dur2022}. The Chinese Parallel mechanism groups choices into bands and finalizes assignments within each band before considering the next \citep{chen2017}. Yet these mechanisms can undermine their goal of rewarding students' eagerness by encouraging strategic reporting. A student may improve her chance of admission to a safer school by ranking it above a preferred but more competitive school. As a result, a higher reported rank need not reflect a greater degree of genuine eagerness, and the mechanism may reward strategic reporting rather than the eagerness that the policy intends to value. \par
This observation raises the central question of this paper: are the strategic incentives observed in practice a consequence of the particular design of the mechanism, or are they inherent in the policy objective of rewarding eagerness? I model this environment by allowing the central mechanism to construct modified priorities from the original priorities, representing exogenous evaluation of students, and students' submitted preferences. Rank monotonicity requires the modification to reward eagerness in a disciplined manner. If a student has higher original priority than another student and also ranks the school at least as highly, their relative priority should not be reversed. RD stability then requires the resulting assignment to be stable with respect to the modified priorities. These two requirements are jointly essential to the policy objective studied here. Without rank monotonicity, the modification need not meaningfully reflect students' eagerness. Without RD stability, the priority modification need not have any force in the assignment. \par
My main result shows that this tension is fundamental. In every market with strict original priorities, any mechanism that satisfies rank monotonicity and RD stability, and strategy-proofness, is outcome equivalent to the student-proposing Deferred Acceptance mechanism using the original priorities. This result holds in a framework where the designer can choose both the priority modification and matching rules. It permits priority adjustments to depend on the entire preference profile, including modifications that cannot be represented by score deductions. Allowing this flexibility therefore does not resolve the conflict between rewarding students' eagerness and guaranteeing truthful reporting. The result extends to settings in which some or all exogenous priority criteria are treated as less important than students' eagerness. In this setting, stability is imposed on the refined modified priorities obtained by resolving remaining ties with the secondary exogenous criteria. Together with the corresponding rank monotonicity condition and strategy-proofness, the result characterizes the student-proposing DA solely using exogenous priority criteria as the unique mechanism (up to outcome equivalence). \par
The nature of the reporting incentives associated with this incompatibility raises a further concern. The proof highlights the familiar reporting motive of omitting more preferred but selective schools to concentrate reported eagerness on a target school. This motive thus remains relevant to the broad class of rank-dependent mechanisms where priority rewards can more flexibly depend on preference than those observed in existing mechanisms. As pointed out in \citet{pathak2008leveling} and \citet{dur2022}, priority rewards may reflect strategic sophistication rather than students' genuine eagerness and mechanisms may work in favor of more informed and sophisticated students. Therefore, these findings suggest that policymakers should refrain from attempting to reward students' eagerness through rank-dependent priorities. 

\paragraph{Related Literature.}
This paper contributes to the literature on centralized student assignment, including \citet{balinski1999tale} and \citet{abdulkadiroglu2003}. In the standard school-choice model, schools are endowed with exogenous priorities, and the student-proposing Deferred Acceptance mechanism provides a solution that is both strategy-proof and stable \citep{gale1962,dubins1981machiavelli,roth1982economics}. A large subsequent literature studies the properties and limitations of this mechanism. \citet{ergin2002}, \citet{abdulkadirouglu2009strategy}, and \citet{kesten2010} examine the tension between respecting priorities and improving student welfare, while \citet{kojima2010axioms} and \citet{morrill2013alternative} characterize the student-proposing Deferred Acceptance mechanism through properties of matching rules. This paper instead focuses on the role of the priorities themselves and studies what can be achieved when the priorities used for assignment are allowed to depend on students' submitted preferences. \par 
A large literature studies school choice mechanisms in which students' reported preferences affect their admission priorities. For the Boston mechanism, \citet{ergin2006games} study the induced preference revelation game, while \citet{pathak2008leveling} examine the different incentives faced by sincere and sophisticated students. \citet{chen2017} theoretically analyze the Chinese Parallel Mechanism, which is a hybrid of the Boston mechanism and the student-proposing DA, and \citet{chen2019chinese} study its incentive and stability properties experimentally. Several papers also use Chinese college admission reforms to study their effects on student behavior and matching outcomes \citep{bo2019admission,cao2020centralized,ha2020college,chen2020empirical}. \citet{dur2022} analyze the Taiwan assignment mechanism used in high school admission and show how rank-based score deductions create strategic dilemmas for students. These studies document both the practical use of rank-sensitive admission rules and the strategic incentives they generate. This paper complements these findings by asking whether such incentives are specific to the particular design of the mechanisms studied in the literature or are inherent in the underlying objective of rewarding students' eagerness. \par
\citet{sasaki2025} studies school choice with rank-dependent priorities and establishes an incompatibility between stability and strategy-proofness. In his model, each school independently determines how its priority order depends on students' submitted preferences, and the central matching mechanism takes these school-specific priority modification structures as given when determining the assignment. \citet{ayoade2023school} study a specific class of school choice mechanisms that use both school priorities and students' preference ranks. Their Preference Rank Partitioned rules can be represented as DA with preference-dependent modified priorities, and they show that the student-proposing DA is the only strategy-proof rank-partition stable rule. This paper instead considers a central mechanism designer jointly specifying how reported preferences affect schools' priorities as well as the matching rule, while not imposing a particular structure on both modification and matching rules. These differences in formulation allow us to study without any additional assumptions whether a central designer can incorporate students' eagerness while guaranteeing truthful reporting. \par
Another related literature studies the welfare gains of using students' reported preferences to elicit information about underlying cardinal preferences. \citet{abdulkadirouglu2011} show that the Boston mechanism can improve ex-ante welfare compared to the student-proposing DA by allowing students' strategic choices to reveal information about their preference intensities. \citet{troyan2012comparing} studies related welfare comparisons across school choice mechanisms, while \citet{abdulkadirouglu2015expanding} propose a mechanism that allows students to influence how priority ties are resolved in order to incorporate information about preference intensities. These papers highlight the potential welfare gains from allowing students' reported preferences to influence the assignment process. This paper takes a different perspective by keeping the planner's objective entirely ordinal. A higher rank is interpreted directly as greater eagerness rather than as a strategic signal of an underlying cardinal preference. While strategic reporting may reveal additional information about preference intensity, it also requires students to understand and adapt to the strategic environment. Differences in such information and sophistication can therefore translate into differences in assignment outcomes even when students have identical underlying cardinal preferences, creating a fairness concern that may offset the benefit of eliciting cardinal information. The present paper asks whether the planner can instead reward eagerness while preserving truthful reporting, which guarantees another notion of fairness. 

\paragraph{} The rest of the paper is organized as follows. \Cref{sec:model} introduces the model and defines the key axioms. \Cref{sec:main} presents the main result, characterizing the student-proposing DA as the unique mechanism (up to outcome equivalence) satisfying the key axioms. \Cref{sec:analysis} analyzes existing mechanisms and shows how mechanisms observed in practice align and do not align with the axioms, while discussing the importance of rank monotonicity. \Cref{sec:discussion} generalizes the model and shows the main result indeed stands. \Cref{sec:conclusion} concludes. \Cref{app:proofs} contains all the proofs. 

\section{Model}\label{sec:model}
\subsection{Primitives}
Consider a centralized matching market with a finite set of students $I$ and a finite set of schools $C$. Each student $i\in I$ has a strict preference relation $P_i$ over $C\cup\{\emptyset\}$, where $\emptyset$ denotes being unmatched. Let $R_i$ represent its weak counterpart, that is, $x\,R_i\,y$ reads $x\,P_i\,y$ or $x=y$. $P:=(P_i)_{i\in I}$ denotes the preference profile. $P_{-i}:=(P_j)_{j\in I\setminus\{i\}}$ denotes the preference profile of students other than $i$. For each $i\in I$ with $P_i$ and each $c\in C\cup\{\emptyset\}$, let $r_{ic}(P_i)\in\mathbb{N}$ be the rank of $c$ in student $i$'s preference $P_i$. Denote by $\mathcal{P}$ the set of strict preference relations over $C\cup\{\emptyset\}$. Each school $c\in C$ has an original priority order $\succ_c$ over $I$ and a finite capacity $q_c\in\mathbb{N}$. The original priority represents an exogenous evaluation of students not affected by students' preference, such as test scores or GPA. $\succ:=(\succ_c)_{c\in C}$ denotes the original priority profile. Let $\Pi$ be the set of strict original priority orders over $I$. Schools are not strategic agents and prefer accepting any students to keeping their seats empty. \par
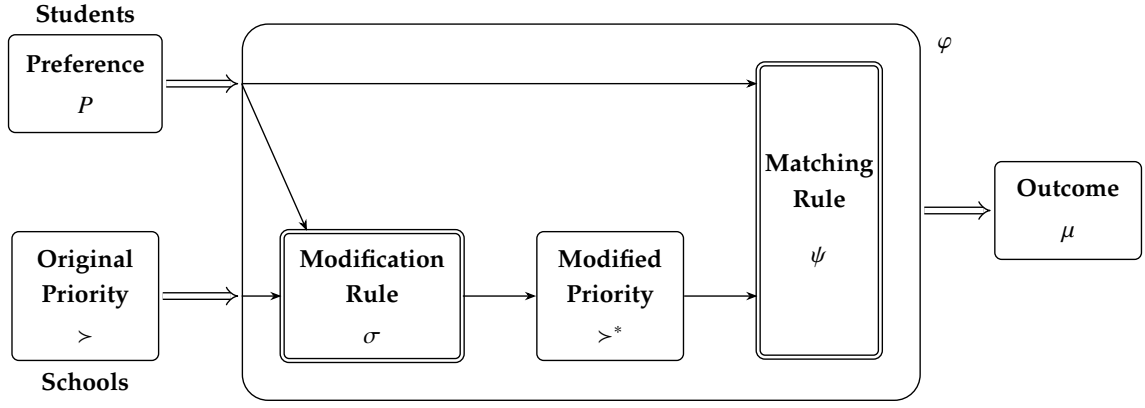
\begin{figure}
    \begin{center}
        % Scales the diagram to fit the width of the slide
        \resizebox{0.95\textwidth}{!}{
        \begin{tikzpicture}[
            node distance=1.5cm and 1cm,
            box/.style={
                rectangle,
                draw=black,
                rounded corners,
                thick,
                minimum width=3cm,
                minimum height=2cm,
                align=center, 
                inner sep=10pt,
                font=\Large\bfseries
            },
            tallbox/.style={
                rectangle,
                draw=black,
                rounded corners,
                thick,
                minimum width=2.5cm,
                minimum height=6cm,
                align=center,
                font=\Large\bfseries
            },
            arrow/.style={
                ->,
                >=Stealth,
                thick
            },
            doublearrow/.style={
                -{Implies},
                double,
                double distance=3pt,
                thick,
                shorten >= 2pt,
                shorten <= 2pt
            }
        ]

            % --- Nodes ---
            
            % Original Priority Box (Used as baseline)
            \node[box] (orig) {Original\\[0.1cm]Priority\\[0.2cm] \normalfont $\succ$};
            \node[font=\Large\bfseries, below=0.1cm of orig] {Schools};

            % Preference Box
            \node[box, above=2cm of orig] (pref) {Preference\\[0.2cm] \normalfont $P$};
            \node[font=\Large\bfseries, above=0.1cm of pref] {Students};

            % Modification Rule Box (Double-lined)
            % Shifted to right=2.5cm to give the input double arrows a medium length
            \node[box, double, double distance=1.5pt, right=2.5cm of orig] (modrule) {Modification\\[0.1cm]Rule\\[0.2cm] \normalfont $\sigma$};

            % Modified Priority Box
            \node[box, right=1.5cm of modrule] (modprio) {Modified\\[0.1cm]Priority\\[0.2cm] \normalfont $\succ^*$};

            % Matching Rule Tall Box (Double-lined)
            % Shifted up so it horizontally aligns with the incoming preference arrow
            \node[tallbox, double, double distance=1.5pt, right=1.5cm of modprio, yshift=1.75cm] (match) {Matching\\[0.1cm]Rule\\[0.5cm] \normalfont $\psi$};

            % The Large Bounding Box (Mechanism Container)
            \node[draw=black, thick, rounded corners=15pt, fit=(modrule) (modprio) (match), inner sep=0.8cm] (mech) {};
            \node[anchor=west, font=\Large] at ($(mech.north east)+(0.2cm,-0.45cm)$) {$\varphi$};

            % Outcome Box
            \node[box, right=1.5cm of mech.east |- match.center] (outcome) {Outcome\\[0.2cm] \normalfont $\mu$};

            % --- Arrows ---
            
            % 1. Find intersection points on the large box boundary
            \coordinate (in_pref) at (mech.west |- pref.east);
            \coordinate (in_orig) at (mech.west |- orig.east);
            
            % 2. Input double arrows from Students/Schools to the large box
            \draw[doublearrow] (pref.east) -- (in_pref);
            \draw[doublearrow] (orig.east) -- (in_orig);
            
            % 3. Internal single arrows starting from the large box boundary
            \draw[arrow] (in_pref) -- (match.west |- in_pref); % Straight to matching rule
            \draw[arrow] (in_pref) -- (modrule.135);           % Diagonal down to modification rule
            \draw[arrow] (in_orig) -- (modrule.west |- in_orig); % Straight to modification rule
            
            % 4. Internal single arrows between the components
            \draw[arrow] (modrule.east) -- (modprio.west);
            \draw[arrow] (modprio.east) -- (match.west |- modprio.east);

            % 5. Output double arrow from the large box boundary to the Outcome
            \coordinate (out_match) at (mech.east |- match.center);
            \draw[doublearrow] (out_match) -- (outcome.west |- match.center);

        \end{tikzpicture}
        }
    \end{center}
    \caption{A mechanism $\varphi$ consists of a modification rule $\sigma$ and a matching rule $\psi$. The modification rule $\sigma$ takes the original priority profile $\succ$ and the preference profile $P$ as inputs and outputs a modified priority profile $\succ^*$. The matching rule $\psi$ takes the modified priority profile $\succ^*$ and the preference profile $P$ as inputs and outputs an outcome $\mu$.}
    \label{fig:mechanism}
\end{figure}

Given a market $\left(I,C,(q_c)_{c\in C}\right)$, a mechanism $\varphi:\mathcal{P}^{|I|}\times\Pi^{|C|}\to\mathcal{M}$, where $\mathcal{M}$ is the set of feasible matchings $\mu:I\to C\cup\{\emptyset\}$, consists of a modification rule $\sigma$ and a matching rule $\psi$, which are summarized in \Cref{fig:mechanism} and detailed below.
Given a preference profile $P\in\mathcal{P}^{|I|}$ and an original priority profile $\succ\in\Pi^{|C|}$, the modification rule $\sigma:\mathcal{P}^{|I|}\times\Pi^{|C|}\to\Pi^{|C|}$ constructs a modified priority profile, which incorporates the students' preference into exogenous evaluation of students. Then, using this newly constructed modified priority profile as well as the preference profile, the matching rule $\psi: \mathcal{P}^{|I|}\times\Pi^{|C|}\to\mathcal{M}$ outputs the resulting matching. Let $\varphi(P,\succ)$ denote the matching under the mechanism $\varphi$ with preference profile $P$ and original priority profile $\succ$. Especially, denote by $\varphi_i(P,\succ)\in C\cup\{\emptyset\}$ and $\varphi_c(P,\succ)\subset I$ student $i$'s assignment and school $c$'s assignment, respectively. 

\subsection{Objectives}
The planner's objective is to design a mechanism that incorporates students' reported preference rankings into school priorities and the resulting assignment while preserving truthful reporting. Rank monotonicity restricts how preference reports modify the original priorities, while RD stability requires the assignment to be stable with respect to the modified priorities. Strategy-proofness ensures that no student can obtain a better outcome by misreporting her preferences regardless of what other students report. The following definitions formalize these three properties. \par
\begin{definition}[Strategy-proofness]
    A mechanism $\varphi$ is strategy-proof if for any $i\in I$, $\succ\in \Pi^{|C|}$, $P\in\mathcal{P}^{|I|}$, and $P'_i\in\mathcal{P}$, 
    \begin{align*}
        \varphi_i(P,\succ)\; R_i\; \varphi_i((P'_i,P_{-i}),\succ).
    \end{align*}
\end{definition}
Since I assume that students are the only strategic agents, I restrict attention to the incentive constraint of students. Observe that the restriction is on the whole mechanism $\varphi$, not on the matching rule $\psi$. A change in a student's report affects both how priorities are modified and the resulting matching. Strategy-proofness of the matching rule for fixed priorities does not by itself guarantee strategy-proofness of the overall mechanism. \par
Rank monotonicity works as a fairness condition for the modification rule. This requires the modification to preserve the relative priority whenever a student with higher original priority ranks a school weakly higher than another student.
\begin{definition}[Rank monotonicity]
    A mechanism $\varphi=(\sigma,\psi)$ satisfies rank monotonicity if for any $P\in\mathcal{P}^{|I|}$, $\succ\in\Pi^{|C|}$, $c\in C$, and $i,j\in I$ such that $i\succ_c j$, $r_{ic}(P_i)\le r_{jc}(P_j)$ implies $i\succ^*_c j$, where $\succ^*=\sigma(P,\succ)$.
\end{definition}
The relative order can be reversed when original priority and preference rank conflict. But this does not mean the relative order needs to be swapped whenever such a conflict happens. The mechanism designer can choose to modify such priorities or leave them unchanged. In particular, a modification rule which leaves the original priority unchanged is rank monotone. \par
RD stability is a fairness condition for the matching rule. This requires the matching rule to be stable with respect to the modified priority as well as the reported preference. To formally define this condition, I first introduce the standard notion of stability of a matching.
\begin{definition}[Stability]
    A matching $\mu$ is stable if
    \begin{itemize}
        \item individual rationality: for any $i\in I$, $\mu(i)\;P_i\;\emptyset$ or $\mu(i)=\emptyset$;
        \item non-wastefulness: for any $i\in I$ and $c\in C$, $c\;P_i\;\mu(i)$ implies $|\mu(c)|=q_c$; and
        \item no blocking pair: there exists no pair $(i,c)\in I\times C$ such that $c\;P_i\;\mu(i)$ and $i\succ_c j$ for some $j\in\mu(c)$.
    \end{itemize}
\end{definition}
Using this, RD stability is defined as follows. 
\begin{definition}[RD stability]
    A mechanism $\varphi=(\sigma,\psi)$ is RD (rank dependent) stable if for any $P\in\mathcal{P}^{|I|}$ and $\succ\in\Pi^{|C|}$, $\varphi(P,\succ)$ is stable with respect to $(P,\succ^*)$, where $\succ^*=\sigma(P,\succ)$. 
\end{definition}
In standard school choice literature, stability is evaluated with respect to the exogenous evaluations, such as test scores and GPA. However, in a rank-dependent context, the modified priority reflects the planner's intended treatment of students based not only on their exogenous evaluation but also on their reported preferences. When a reversal occurs from the original priority, an assignment that respects the modified priority may create justified envy with respect to the original priority. Evaluating stability based on the original priority would then reject an assignment that is consistent with the planner's intended priority modification. Thus, I treat the modified priority as the relevant priority for stability, reflecting the rank-dependent aspect of school priorities. \par
Taken together, the axioms establish the normative framework for a rank-dependent mechanism. Especially, rank monotonicity and RD stability jointly capture the policy goal of rewarding students' eagerness. Without rank monotonicity, RD stability alone puts no restriction on how modified priorities are constructed. Without RD stability, rank monotonicity alone does not regulate how the modified priorities are used in the assignment. I seek a mechanism that allows schools to prioritize eager students without acting arbitrarily (rank monotonicity), while maintaining the standard guarantees of truthful reporting (strategy-proofness) and fairness (RD stability). 

\section{Uniqueness of the Student-Proposing DA}\label{sec:main}
The framework established in the preceding section allows the designer to choose both the priority modification rule and the matching rule. The question is whether this flexibility can be used to design a mechanism that rewards students' eagerness while guaranteeing truthful reporting. The following result negatively answers this question by characterizing the student-proposing DA mechanism as the unique mechanism (up to outcome equivalence) satisfying strategy-proofness, rank monotonicity, and RD stability.
%The preceding section shows the limitations of existing mechanisms in implementing the policy goal of rewarding students' eagerness while maintaining truthful reporting. However, the framework still leaves the designer substantial flexibility in choosing the priority modification rule and the matching rule. The following result answers the remaining question of whether this freedom can be used to design a strategy-proof mechanism that satisfies RD stability and rank monotonicity. \memo{Revise the opening.}
\begin{theorem}\label{thm:noexist}
    For any market $\left(I,C,(q_c)_{c\in C}\right)$, a mechanism that satisfies strategy-proofness, RD stability, and rank monotonicity is outcome equivalent to the student-proposing DA. 
\end{theorem}
The proof is relegated to \Cref{proof:noexist}. Here I show its intuition. Suppose toward contradiction that a strategy-proof mechanism satisfying rank monotonicity and RD stability differs from the student-proposing DA. The result of \citet{alva2019} implies that there is a preference profile where a student prefers her DA assignment to the mechanism's assignment. To prevent her from listing only her DA assignment, the mechanism must leave her unmatched under such a preference report. \par
At this new preference profile, RD stability requires this school to be fully filled by students with higher modified priority than her. Since she now ranks the school at the top, rank monotonicity requires that all these students must have higher original priority than her. Here, her DA assignment under the new preference profile does not change. Thus, at least one of these students has a worse assignment than she should get under the student-proposing DA using the new preference profile. The same argument applies to this student, and each of newly worse assigned students must have at least two acceptable schools and thus differs from the students whose report has already been truncated. Repeating the construction would require an endless sequence of students, contradicting the finiteness of the student set. \par
\Cref{thm:noexist} applies to every finite market $\left(I,C,(q_c)_{c\in C}\right)$ and strict original priority $\succ$, without any restriction on school capacities or the structure of strict original priorities. In each of these markets, any mechanism satisfying strategy-proofness, rank monotonicity, and RD stability must reproduce the student-proposing DA assignment under the original priority for every preference profile. At the same time, the student-proposing DA can be interpreted as a mechanism that has a modification rule that leaves the original priority unchanged and a matching rule of $DA^S$. Thus, within this framework, a mechanism cannot reward students' eagerness by departing from the student-proposing DA assignment while maintaining truthful reporting. Modified priorities may still depend on reported preferences, but the resulting assignment should not effectively reflect such modification and has to coincide with the standard DA assignment. \par
The characterization shows that considering the class of modifications beyond deduction rules does not resolve the incentive problem. Rank monotonicity allows priority rewards to depend on the entire preference profile.\footnote{See \Cref{ex:rankmonotone} for more detail.} The designer can for instance adjust rewards based on the market demand for each school, without fixing a predetermined bonus for each preference rank. However, \Cref{thm:noexist} shows that even with this flexibility, we cannot depart from the student-proposing DA without violating strategy-proofness. The result is robust to the choice of matching rule as well. Even if we allow the designer to choose a matching rule other than the student-proposing DA, any mechanism that satisfies the three properties must always reproduce the student-proposing DA assignment. \par
The proof of \Cref{thm:noexist} relies on a specific construction of deviation, but it connects this broader impossibility to the motive of concentrating reported eagerness on a target school, which we often see in mechanisms in practice, most notably the Boston mechanism. Under deduction rules, a student can potentially improve her assignment by removing more preferred but selective schools from her preference to avoid a larger deduction at a less preferred but more accessible school. The deviation considered in the proof is an extreme form of this type of manipulation, where a student lists only the school she can secure by appropriately reporting her preference. The proof exposes that this type of relatively simple manipulation we see today is not a product of a simple design, but a property persistent even in the framework that allows for more complex and context-dependent priority modifications. \par
This incentive raises a practical concern for policymakers who seek to reward students' eagerness. For instance, the equilibrium analyses by \citet{pathak2008leveling} for the Boston mechanism and \citet{dur2022} for the Taiwan assignment mechanism point out that strategic sophistication can give students an advantage over those who report sincerely.\footnote{\citet{dur2018identifying} provide empirical evidence from Wake County's Boston-style assignment system. Applicants classified as strategically sophisticated were $9.6$ percentage points more likely to obtain one of their preferred schools, with the advantage associated with systematically avoiding over-demanded schools.} Families may differ in their ability to recognize the over-demanded schools and figure out what kind of strategic promotion of attainable schools improves their assignment. Consequently, priority rewards may not only represent students' genuine eagerness, but also reflect their families' ability to adapt to the market and exploit the mechanism. RD stability does not address this concern since the modified priorities depend on reports that families can adjust strategically. \Cref{thm:noexist} establishes that, under rank monotonicity and RD stability, any mechanism that attempts to deviate from the student-proposing DA to reward students' eagerness will inevitably face an incentive issue that has been observed in practice and concerned to be problematic. Policymakers who agree with the requirement of rank monotonicity and RD stability should therefore retain the student-proposing DA as the only viable solution even if they wish to reward students' eagerness. \par
It is important to note that the framework in this paper applies rank monotonicity to the entire original priority order. However, policymakers may regard some criteria used to construct the original priority as less important than students' eagerness, particularly when those criteria are used mainly to break ties in the original priority. \Cref{sec:discussion} discusses whether allowing eagerness to override such criteria changes the result of \Cref{thm:noexist}.

\section{Analysis of Existing Mechanisms}\label{sec:analysis}
This section discusses how the framework and \Cref{thm:noexist} are related to the existing mechanisms. I first show how the Taiwan assignment mechanism satisfies rank monotonicity and RD stability, and how the framework extends beyond its score-deduction rule. I then show why TTC and EADA cannot be interpreted as mechanisms with a rank monotone modification rule and a matching rule that is stable under modified priority. \par
I first consider the Taiwan assignment mechanism. Formally, this mechanism is defined as a mapping from students' reported preferences and students' exam scores, which is cardinal information, to a matching. For each student $i$ and school $c$ let $s_{ic}\in\mathbb{R}_+$ represent an original score, which is the student's exam score. This represents school $c$'s original priority so that $i\succ_c j$ if and only if $s_{ic}>s_{jc}$. We assume no ties in original scores. \par
The mechanism operates in two stages: score deduction stage and matching stage. In the score deduction stage, each school $c$ has a deduction rule $\lambda_c=(\lambda_{c,1},\lambda_{c,2},\dots,\lambda_{c,|C|+1})\in\mathbb{R}_+^{|C|+1}$ with $\lambda_{c,1}=0$ and $\lambda_{c,k}\le\lambda_{c,k+1}$ for any $k\in\{1,2,\dots,|C|\}$. Then, the modified priority is constructed based on the original score and the deduction rule so that for any $i,j\in I$, $i\succ^*_c j$ if and only if $s_{ic}-\lambda_{c,r_{ic}(P_i)} > s_{jc}-\lambda_{c,r_{jc}(P_j)}$, assuming there is no tie in the modified score. In the matching stage, the student-proposing DA is run using the modified priority $\succ^*$ and the reported preference $P$.\footnote{This representation and assumption is based on \citet{dur2022}.}\par
With appropriate choice of the deduction rule, the Taiwan assignment mechanism can represent the Boston mechanism and the Chinese Parallel mechanism \citep{dur2022}. The Taiwan assignment mechanism is not strategy-proof in general. By removing a selective school from her preference report, a student can reduce the deduction applied to her score at her less preferred school and improve her priority at that school. This may help her get accepted to the school she would be rejected if she wasted her top choices to selective schools she had no chance with. \par
The mechanism nevertheless satisfies RD stability. For every reported preference $P$ and modified priority $\succ^*$, the matching rule $DA^S$ is known to return a stable outcome with respect to $(P,\succ^*)$. Thus, the Taiwan assignment mechanism inherits the fairness aspect of the student-proposing DA, even though its incentive property is in general undermined by the score deduction rule. \par
To see that the mechanism satisfies rank monotonicity, the following result is useful. It shows that the deduction rule is a sufficient condition for rank monotonicity. 
\begin{proposition}\label{prop:sufficient}
    If modified priorities rank students by $s_{ic}-\lambda_{c,r_{ic}(P_i)}$ where $\lambda_{c,k}$ is weakly increasing in $k$ for every school $c$, then the mechanism satisfies rank monotonicity.
\end{proposition}
The proof is relegated to \Cref{proof:sufficient}. A student who ranks a school at least as highly as another student gets weakly smaller deduction. Thus, if she has a higher original score, the score adjustment preserves her priority relative to the other student. However, a rank monotone modification rule permits a more general class of priority modifications than those induced by the deduction rule. In particular, the relative priority of two students may depend on the entire preference profile. The following example illustrates how their relative priority can be affected by the change of the other students' preference reports even when their own preferences are unchanged.
\begin{example}\label{ex:rankmonotone}
    Consider a school $c$ and three students with original priority
    \begin{align*}
        1\succ_c 2\succ_c 3.
    \end{align*}
    Suppose that there are two other schools. School $c$ can be ranked first, second, or third. Consider the following modification rule. If exactly one student ranks $c$ first, that student is given higher modified priority than every other student, while the original priority is preserved among the remaining students. If either no student or at least two students rank $c$ first, the original priority is unchanged.
\end{example}
This modification rule satisfies rank monotonicity. Consider any $i,j\in I$ such that $i\succ_c j$ and $r_{ic}\le r_{jc}$. If the original priority is unchanged, then $i\succ_c^*j$. When exactly one student ranks $c$ first, if $i$ is that student, then $i\succ_c^*j$ by construction. If neither $i$ nor $j$ is that student, their original priority is preserved. Finally, $j$ cannot be the unique student ranking $c$ first, since this would imply $r_{jc}=1<r_{ic}$, contradicting $r_{ic}\le r_{jc}$. Hence, $i\succ_c^*j$ in every case. \par
However, this modification cannot be represented by the deduction rule described above. Consider two preference profiles $P$ and $P'$ such that
\begin{align*}
    r_{1c}(P_1)=r_{1c}(P'_1)=2,\qquad r_{2c}(P_2)=r_{2c}(P'_2)=1,\qquad r_{3c}(P_3)=3,\qquad r_{3c}(P'_3)=1.
\end{align*}
At $P$, student $2$ is the unique student who ranks $c$ first, so $2\succ_c^*1$. At $P'$, both students $2$ and $3$ rank $c$ first, so no priority benefit is given, and the original priority is restored: $1\succ_c^*2$. Thus, the relative modified priority between students $1$ and $2$ changes even though neither student's rank of $c$ changes. Under the deduction rule, however, their relative modified priority is determined solely by $s_{1c}-\lambda_{c,2}$ and $s_{2c}-\lambda_{c,1}$. Since these values are unchanged from $P$ to $P'$, their relative priority cannot change. Therefore, no deduction rule can represent this rank monotone modification rule.\footnote{Note that \Cref{ex:rankmonotone} cannot be represented by a student-specific deduction rule as well, where the amount of deduction can be different for each student.} \par
This example shows that rank monotonicity allows a planner to condition the reward for eagerness on the market demand for a school. If a student ranks a low demand school highly, she may be rewarded. If a school is highly demanded, the planner may choose to not reward students' eagerness even if they rank the school highly. This is one form of additional flexibility covered by \Cref{thm:noexist}, and the characterization under this broader framework therefore addresses the compatibility of the policy goal with truthful reporting without relying on a particular structure of priority adjustments. \par 
Next, I consider mechanisms that pursue student-side efficiency, such as TTC and EADA proposed by \citet{kesten2010}. These mechanisms and rank dependent mechanisms both aim to make assignments more reflective of students' preferences. However, the priority reversals used to improve students' welfare need not be rank monotone. Since TTC is strategy-proof and can differ from the student-proposing DA, \Cref{thm:noexist} rules out an interpretation of TTC as a mechanism satisfying rank monotonicity and RD stability. EADA can Pareto improve on the student-proposing DA while sacrificing strategy-proofness, but these welfare gains need not be consistent with the policy goal of rewarding students' eagerness. The following example illustrates that both TTC and EADA can produce an outcome that no rank monotone modification can make RD stable. 
\begin{example}\label{ex:ttc}
Consider a setting with $I=\{1,2,3,4\}$, $C=\{a,b,c,d\}$, and $q_a=q_b=q_c=q_d=1$. The preference and original priority are given as:
\begin{align*}
    &P_1:a,b,c,d\qquad\succ_a:1,2,3,4\\
    &P_2:a,b,c,d\qquad\succ_b:3,4,2,1\\
    &P_3:c,b,a,d\qquad\succ_c:2,3,4,1\\
    &P_4:b,d,a,c\qquad\succ_d:4,1,2,3
\end{align*}
\end{example}
The outcome of TTC is given by $TTC(P,\succ)=((1,a),(2,b),(3,c),(4,d))$. There exists no rank monotone modification that makes this outcome RD stable. To see this, consider a tuple $(2,4,b)$. Since $b\,P_4\,d$, RD stability requires $2\succ_b^*4$. However, since $4\succ_b2$ and $r_{2b}(P_2)=2>r_{4b}(P_4)=1$, rank monotonicity requires $4\succ_b^*2$. \par
EADA is known to be Pareto efficient when all students consent to priority waivers while relaxing the requirement of stability. Even though the approach of relaxing stability with respect to the original priorities is common to our approach, EADA does not jointly satisfy rank monotonicity and RD stability. To see this, consider the same example as in \Cref{ex:ttc}. The outcome of EADA is also given by $EADA(P,\succ)=((1,a),(2,b),(3,c),(4,d))$. As before, no rank monotone modification makes this outcome RD stable. \par
Nevertheless, the assignment $((1,a),(2,b),(3,c),(4,d))$ can be obtained by applying the student-proposing DA to suitably modified priorities. Keep the original priority except for school $b$, and let $3\succ_b^*2\succ_b^*4\succ_b^*1$. Then, the student-proposing DA with this modified priority returns the same outcome as $TTC(P,\succ)=EADA(P,\succ)$. \par
This example illustrates that reproducing an assignment using modified priorities does not by itself establish that these mechanisms respect the policy goal of rewarding students' eagerness. A mechanism's consistency with the policy goal depends on how the modified priorities are related to the original priorities and students' preferences. The following result clarifies the role of rank monotonicity. It shows that without this restriction, any Pareto efficient mechanism can admit an outcome equivalent representation using modified priorities and the student-proposing DA. \par
\begin{proposition}\label{prop:sp-efficient}
    For any Pareto efficient mechanism $\nu$, there exists an outcome equivalent mechanism $\varphi^\nu=(\sigma^\nu,DA^S)$.
\end{proposition}
The proof is relegated to \Cref{proof:sp-efficient}. Given an assignment selected by a Pareto efficient mechanism, let each school's assigned students be prioritized over all the other students. The assignment is then stable under the modified priority. Since the student-proposing DA weakly Pareto dominates any other stable assignment under the constructed modified priority, Pareto efficiency ensures that the student-proposing DA recovers the selected assignment. Thus, this construction can support any Pareto efficient assignment if we do not require the modified priority to reflect the policy goal of rewarding students' eagerness. \par
The interpretation of priorities in TTC and EADA illustrates this distinction. Under TTC, priorities provide opportunities for students to obtain preferred schools through trading cycles \citep{abdulkadiroglu2003}. EADA uses consent to waive certain priorities without changing the consenting students' own assignment, potentially enabling improvements for other students \citep{kesten2010}. On the other hand, under the policy goal studied in this paper, fairness constrains how the original priority can be modified to reward students' eagerness. An original priority relation may be reversed only when the lower-priority student ranks the school more highly. Rank monotonicity formalizes this restriction, while RD stability ensures that the assignment respects the modified priority. The failure of TTC and EADA demonstrated in \Cref{ex:ttc} comes from these priority reversals prohibited by rank monotonicity. \Cref{prop:sp-efficient} highlights why a representation using modified priorities and the student-proposing DA is not sufficient and a rank dependent mechanism must be accompanied by a restriction on how those priorities are constructed to capture the policy goal of rewarding students' eagerness. 

\section{When Some Features are Less Important Than Eagerness}\label{sec:discussion}
This section analyzes the case where schools may use secondary criteria to order students who are equally ranked under their main admission criteria. For example, schools may distinguish students with the same total test score by their individual subject scores. In this case, a policymaker aiming to reward students' eagerness may regard this secondary comparison as less important than students' preference ranks. This motivates a formulation in which (weak) original priorities are modified before the secondary criteria are applied to resolve the remaining incomparabilities. I then examine whether this formulation permits a strategy-proof mechanism different from the student-proposing DA. \par
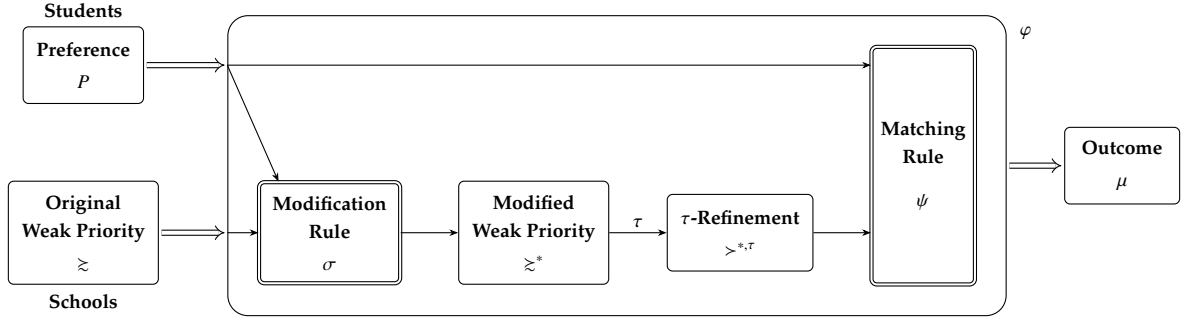
\begin{figure}
    \begin{center}
        \resizebox{0.98\textwidth}{!}{
        \begin{tikzpicture}[
            node distance=1.5cm and 1cm,
            box/.style={
                rectangle,
                draw=black,
                rounded corners,
                thick,
                minimum width=3cm,
                minimum height=2cm,
                align=center,
                inner sep=10pt,
                font=\Large\bfseries
            },
            tallbox/.style={
                rectangle,
                draw=black,
                rounded corners,
                thick,
                minimum width=2.7cm,
                minimum height=6.2cm,
                align=center,
                font=\Large\bfseries
            },
            arrow/.style={
                ->,
                >=Stealth,
                thick
            },
            doublearrow/.style={
                -{Implies},
                double,
                double distance=3pt,
                thick,
                shorten >= 2pt,
                shorten <= 2pt
            }
        ]

            % --- Nodes ---

            % Original weak priority
            \node[box] (orig) {Original\\[0.1cm]Weak Priority\\[0.2cm] \normalfont $\succsim$};
            \node[font=\Large\bfseries, below=0.1cm of orig] {Schools};

            % Preference
            \node[box, above=2cm of orig] (pref) {Preference\\[0.2cm] \normalfont $P$};
            \node[font=\Large\bfseries, above=0.1cm of pref] {Students};

            % Modification rule
            \node[box, double, double distance=1.5pt, right=2.6cm of orig] (modrule) {Modification\\[0.1cm]Rule\\[0.2cm] \normalfont $\sigma$};

            % Modified weak priority
            \node[box, right=1.5cm of modrule] (modprio) {Modified\\[0.1cm]Weak Priority\\[0.2cm] \normalfont $\succsim^*$};

            % Tau-refinement
            \node[box, right=1.5cm of modprio] (refine) {$\tau$-Refinement\\[0.2cm] \normalfont $\succ^{*,\tau}$};

            % Matching rule
            \node[tallbox, double, double distance=1.5pt, right=1.5cm of refine, yshift=1.75cm] (match) {Matching\\[0.1cm]Rule\\[0.5cm] \normalfont $\psi$};

            % Mechanism container
            \node[draw=black, thick, rounded corners=15pt, fit=(modrule) (modprio) (refine) (match), inner sep=0.8cm] (mech) {};
            \node[anchor=west, font=\Large] at ($(mech.north east)+(0.2cm,-0.45cm)$) {$\varphi$};

            % Outcome
            \node[box, right=1.5cm of mech.east |- match.center] (outcome) {Outcome\\[0.2cm] \normalfont $\mu$};

            % --- Arrows ---

            % Boundary coordinates
            \coordinate (in_pref) at (mech.west |- pref.east);
            \coordinate (in_orig) at (mech.west |- orig.east);

            % Inputs to mechanism
            \draw[doublearrow] (pref.east) -- (in_pref);
            \draw[doublearrow] (orig.east) -- (in_orig);

            % Internal arrows from boundary
            \draw[arrow] (in_pref) -- (match.west |- in_pref);
            \draw[arrow] (in_pref) -- (modrule.135);
            \draw[arrow] (in_orig) -- (modrule.west |- in_orig);

            % Internal arrows between components
            \draw[arrow] (modrule.east) -- (modprio.west);

            \draw[arrow] (modprio.east) -- node[above, font=\Large] {$\tau$} (refine.west);

            \draw[arrow] (refine.east) -- (match.west |- refine.east);

            % Output
            \coordinate (out_match) at (mech.east |- match.center);
            \draw[doublearrow] (out_match) -- (outcome.west |- match.center);

        \end{tikzpicture}
        }
    \end{center}
    \caption{A mechanism now allows a weak original priority $\succsim$ endowed with a tie-breaking rule $\tau$. The modification rule $\sigma$ outputs a modified priority $\succsim^*$, which is allowed to be weak. The matching rule $\psi$ takes the $\tau$-refinement of the modified priority $\succsim^{*,\tau}$ as well as preference $P$ and outputs the outcome $\mu$.}
    \label{fig:mechanism-weak}
\end{figure}

In this section, the framework is extended to allow schools to have weak original priority $\succsim$, which represents the main admission criteria, as well as a strict refinement rule $\tau$, which represents the secondary criteria. Then, a mechanism receives preference $P$, weak original priority $\succsim$, and the refinement rule $\tau$, and outputs the outcome $\mu=\varphi(P,\succsim,\tau)$. Specifically, modification rule receives the weak original priority $\succsim$ and reported preference $P$ and outputs (potentially) weak modified priority $\succsim^*$. Then, the matching rule receives $\succsim^{*\tau}$, which is a $\tau$-refinement of $\succsim^*$, as well as preference $P$ and outputs the matching $\mu$. See \Cref{fig:mechanism-weak} for the graphical representation of the mechanism. 
To accommodate incomparability in the original priority, I adapt the definition of rank monotonicity. The revised definition preserves a strict priority advantage when it is consistent with students' preference ranks. When students are originally tied, a student who ranks the school less highly should not receive a strict advantage. The following definition formalizes these requirements.
\begin{definition}[Weak rank monotonicity]
    A mechanism $\varphi=(\sigma,\psi)$ satisfies weak rank monotonicity if for any $P$, $\succsim$, $\tau$, $c\in C$, and $i,j\in I$, 
    \begin{align*}
        i\succ_cj,\;r_{ic}(P_i)\le r_{jc}(P_j)\quad &\Rightarrow\quad i\succ^*_cj,\\
        i\succsim_cj,\;r_{ic}(P_i)\le r_{jc}(P_j)\quad &\Rightarrow\quad i\succsim^*_cj,
    \end{align*}
    where $\succsim^*=\sigma(P,\succsim)$.
\end{definition}
This definition requires students who are tied in the original priority and rank a school the same to be kept tied. When their preference ranks differ while tied in the original priority, their relative priority order may either be preserved as a tie or work in favor of the student who ranks the school more highly. \par
Any ties remaining after the priority modification are resolved by the secondary criteria $\tau$. The priority comparisons newly induced by $\tau$ are relevant for fairness, contrary to the standard tie-breaking rule considered in school choice literature with weak priorities. I therefore require the outcome to respect the modified priorities after the refinement. The following definition formalizes this requirement.
\begin{definition}[Strong RD stability]
    A mechanism $\varphi=(\sigma,\psi)$ satisfies strong RD stability if for any $P$, $\succsim$, and $\tau$, the outcome $\mu=\varphi(P,\succsim,\tau)$ is stable with respect to $(P,\succsim^{*,\tau})$.
\end{definition}
The assumption of referring to the refined priorities is relevant in practice when ties in a coarse priority are resolved using additional criteria. For example, Chinese college admission systems often use subject-specific examination scores to rank students with the same aggregate score.\footnote{For example, the 2026 admission rules of Fujian Province specify that applicants with the same total Gaokao score are ranked successively by the combined Chinese and mathematics score, the higher of the two subject scores, foreign language score, the primary elective subject score, the highest secondary elective subject score, and so on. If all criteria are exhausted and students remain tied, the school chooses which students to admit. See Fujian Provincial Department of Education, \href{https://jyt.fujian.gov.cn/xxgk/zywj/202606/t20260612_7162702.htm}{2026 Fujian Province General Higher Education Admission Implementation Measures}.} Such refinements differ from arbitrary tie-breaking because they reflect criteria that the admission authority explicitly uses to determine which students to be prioritized. It is therefore natural to treat the resulting strict ranking as the relevant priority for stability. Moreover, when students remain tied after all prescribed criteria are applied, Taiwanese high-school admission systems increase the number of admitted students rather than introducing an additional arbitrary tie-breaking comparison.\footnote{See the Taipei City 12-Year Basic Education Information Network, \href{https://12basic.tp.edu.tw/faq/faq_01/\%E4\%BD\%95\%E8\%AC\%82\%E8\%B6\%85\%E9\%A1\%8D\%E6\%AF\%94\%E5\%BA\%8F/}{``What is Comparison Ranking?''} The Q\&A states that if applicants remain tied after all prescribed comparisons, additional places may be approved, generally up to five percent of the school's capacity.} This practice further suggests that the preceding refinements represent meaningful priority comparisons. \par
Under these requirements, the following result shows that \Cref{thm:noexist} extends to the case where schools have weak original priorities and a refinement rule. Strategy-proofness requires the mechanism to be the student-proposing DA applied to the $\tau$-refinement of the original priority.
\begin{corollary}\label{cor:noexist-weak}
    If a mechanism satisfies strategy-proofness, weak rank monotonicity, and strong RD stability, then it is uniquely outcome equivalent to the student-proposing DA with a tie-breaking rule $\tau$.
\end{corollary}
The proof is relegated to \Cref{proof:noexist-weak}. The argument treats the $\tau$-refinements of the original and modified priorities as the strict priorities in the proof of \Cref{thm:noexist}. Consider a student who ranks the school at least as highly as another student and has a higher original priority under the refined original priority. If the priority advantage comes from the original priority, weak rank monotonicity preserves it strictly as in the proof of \Cref{thm:noexist}. If it comes from the refinement, weak rank monotonicity ensures the student is prioritized at least as highly as the other student. Then, any remaining tie is resolved in favor of her by the refinement $\tau$. Thus, the refined priorities satisfy rank monotonicity in the original framework. Since strong RD stability requires stability under the refined modified priorities and strategy-proofness is unchanged, \Cref{thm:noexist} applies to obtain the result. \par
\Cref{cor:noexist-weak} shows that even if we value eagerness over some or all of the relevant criteria, the incentive problem persists. In particular, the result applies even when policymakers preserve all strict comparisons in the main criteria and only allow eagerness to compare tied students in the original priority. The result also covers the opposite extreme where eagerness is treated as more important than all exogenous criteria. This case is represented as a weak original priority that treats all students as tied, a weak modified priority ordering them by how highly they rank each school, and the refinement only distinguishing students reporting the same rank using the exogenous criteria. In either case, if a mechanism satisfies weak rank monotonicity and strong RD stability, strategy-proofness requires the mechanism to reproduce the outcome of the student-proposing DA with the original priority and the refinement rule. 

\section{Conclusion}\label{sec:conclusion}
This paper studies whether school choice mechanisms can reward students' eagerness through rank dependent priorities while guaranteeing truthful reporting. I show that, in every finite market with strict original priorities, any mechanism satisfying strategy-proofness, rank monotonicity, and RD stability must be outcome equivalent to the student-proposing DA. This characterization is obtained in the framework where the designer can choose both how priorities are modified and how students are assigned. The conclusion holds even when some or all exogenous priority criteria are treated as less important than students' eagerness, under the corresponding rank monotonicity condition and stability with respect to the refined priorities. Thus, under the fairness requirements of rank monotonicity and RD stability, it is structurally impossible to implement a mechanism that differs from the student-proposing DA to reward students' eagerness while preserving truthful reporting. \par
A future direction of research may explore mechanisms while relaxing the incentive compatibility requirement. Such a relaxation should be formulated in terms of ordinal preferences and provide a meaningful guarantee under plausible assumptions on students' information and strategic behavior in real-world school choice environments. 

\appendix
\section{Proofs}\label{app:proofs}
\subsection{Proof of \Cref{thm:noexist}\protect\footnote{Note that the proof stands even if we allow the modified priority to be weak without any meaningful refinement supplied.}}\label{proof:noexist}
The student-proposing DA is known to be strategy-proof. Also, the student-proposing DA is represented as a modification rule that always leaves the original priority unchanged and a matching rule of $DA^S$, which implies rank monotonicity and RD stability. To see its uniqueness up to outcome equivalence, I first introduce the following result by \citet{alva2019}.
\begin{lemma}[\citealp{alva2019}, Corollary 5]\label{lemma:alva2019}
   If $\succ$ consists of strict priorities, then there exists no strategy-proof mechanism that Pareto dominates the student-proposing Deferred Acceptance mechanism. 
\end{lemma}
Proof of \Cref{thm:noexist} is by induction and contradiction. Take any market $\left(I, C, (q_c)_{c\in C}, \succ\right)$. By \Cref{lemma:alva2019}, $\varphi$ cannot Pareto improve $DA^S$, that is, there exist $P^1\in\mathcal{P}^{|I|}$ and $i^1\in I$ such that 
$$c^1:=DA^S_{i^1}(P^1,\succ)\,P^1_{i^1}\,\varphi_{i^1}(P^1,\succ).$$ 
Consider a student $i^1$'s deviation of $P^2_{i^1}:c^1,\emptyset$ and let $P^2=(P^2_{i^1},P^1_{-i^1})$. Since $\varphi$ is strategy-proof, it must be that $\varphi_{i^1}(P^2,\succ)=\emptyset$. By RD stability, it must be that $|\varphi_{c^1}(P^2,\succ)|=q_{c^1}$ and $j\succ^*_{c^1}i^1$ for any $j\in\varphi_{c^1}(P^2,\succ)$. Since $r_{i^1c^1}(P^2_{i^1})=1$, rank monotonicity implies $j\succ_{c^1}i^1$ for any $j\in\varphi_{c^1}(P^2,\succ)$. Observe that $DA^S_{i^1}(P^2,\succ)=c^1$ by the rural hospital theorem.\footnote{More specifically, $\mu:=DA^S(P^1,\succ)$ is stable under $(P^2,\succ)$. The rural hospital theorem \citep[Lemma 1]{roth1986allocation} ensures that the set of matched students is the same under all stable matchings, implying that $i^1$'s assignment does not change.} Then, stability of $DA^S$ implies that there exists $i^2\in\varphi_{c^1}(P^2,\succ)$ such that 
$$c^2:=DA^S_{i^2}(P^2,\succ)\,P^2_{i^2}\,\varphi_{i^2}(P^2,\succ)=c^1.$$ 
Now, given students $i^1,\dots,i^{k-1}$, consider student $i^k$ ($k\ge 2$). Observe that $i^k\neq i^l$ for any $l=1,\dots,k-1$. This is because at $P^k$, student $i^l$ only lists school $c^l$ in her preference $P^k_{i^l}$. On the other hand, for student $i^k$, it holds that
$$c^k=DA^S_{i^k}(P^k,\succ)\,P^k_{i^k}\,\varphi_{i^k}(P^k,\succ)=c^{k-1}.$$ 
By individual rationality of $\varphi$, $i^k$ lists at least two schools, $c^k$ and $c^{k-1}$, over the outside option in her preference $P^k_{i^k}$. Now, consider $i^k$'s deviation $P^{k+1}_{i^k}:c^k,\emptyset$ and let $P^{k+1}=(P^{k+1}_{i^k},P^k_{-i^k})$. Strategy-proofness of $\varphi$ implies $\varphi_{i^k}(P^{k+1},\succ)=\emptyset$, and RD stability and rank monotonicity jointly imply $|\varphi_{c^k}(P^{k+1},\succ)|=q_{c^k}$ and $j\succ_{c^k}i^k$ for all $j\in\varphi_{c^k}(P^{k+1},\succ)$. Since $DA^S_{i^k}(P^{k+1},\succ)=c^k$ from the rural hospital theorem, by stability of $DA^S$, there exists $i^{k+1}\in\varphi_{c^k}(P^{k+1},\succ)$ with
$$c^{k+1}:=DA^S_{i^{k+1}}(P^{k+1},\succ)\,P^{k+1}_{i^{k+1}}\,\varphi_{i^{k+1}}(P^{k+1},\succ)=c^k.$$
Thus, we obtain a sequence of $\{i^n,c^n,P^n\}$. However, since $I$ is finite, at some $m\in\mathbb{N}$, 
$$\varphi_{c^m}(P^{m+1},\succ)\setminus\{i^1,\dots,i^{m-1}\}=\emptyset.$$
Then, for all $j\in \varphi_{c^m}(P^{m+1},\succ)$, $DA^S_j(P^{m+1},\succ)=c^m$. Also, $DA^S_{i^m}(P^{m+1},\succ)=c^m$. This requires $|\varphi_{c^m}(P^{m+1},\succ)|\le q_{c^m}-1$. At the same time, by $\varphi$'s strategy-proofness, $\varphi_{i^m}(P^{m+1},\succ)=\emptyset$. This violates the non-wastefulness of $\varphi$, completing the proof. \qed

\subsection{Proof of \Cref{prop:sufficient}}\label{proof:sufficient}
Consider $P$, $\succ$, $i,j\in I$, and $c\in C$ such that $i\succ_c j$ and $r_{ic}(P_i)\le r_{jc}(P_j)$. Then, we have $s_{ic}>s_{jc}$ and $\lambda_{c,r_{ic}(P_i)}\le\lambda_{c,r_{jc}(P_j)}$. Thus, $s_{ic}-\lambda_{c,r_{ic}(P_i)} > s_{jc}-\lambda_{c,r_{jc}(P_j)}$, which implies $i\succ^*_c j$. \qed

\subsection{Proof of \Cref{prop:sp-efficient}}\label{proof:sp-efficient}
I first introduce the following result by \citet{gale1962}. 
\begin{lemma}[\citealp{gale1962}, Theorem 2]\label{lemma:unique}
    If a matching rule $\psi$ is student-optimal, then $\psi$ is uniquely outcome equivalent to the student-proposing DA. 
\end{lemma}
Now, let $\nu$ be a Pareto efficient mechanism. For any student preference profile $P$ and school original priority profile $\succ$, construct a modification rule $\sigma$ so that for any school $c\in C$, students assigned to school $c$ under $\nu(P,\succ)$ are listed at the top of $\succ^*_c$. The relative orders among assigned students and among unassigned students can be arbitrary. \par
Now, I show that $\nu(P,\succ)$ is stable with respect to $(P,\succ^*)$. Since $\nu$ is Pareto efficient, $\nu(P,\succ)$ is RD non-wasteful. Also, under $\succ^*$, every school $c$ is matched with the top candidates in their modified priority order $\succ^*_c$. Thus, no school can form a blocking pair with respect to $\succ^*$. \par
The Pareto efficiency and RD stability of $\nu(P,\succ)$ together imply the RD constrained efficiency\footnote{RD constrained efficiency is defined as a mechanism's outcome is always Pareto undominated by RD stable matchings.} of the mechanism $\nu$. Hence, \Cref{lemma:unique} suggests $\nu(P,\succ)=DA^S(P,\succ^*)$. This implies for any $P\in\mathcal{P}^{|I|}$ and $\succ\in\Pi^{|C|}$, $\varphi=(\sigma,DA^S)$ can recover any mechanism $\nu$ that is Pareto efficient. \qed

\subsection{Proof of \Cref{cor:noexist-weak}}\label{proof:noexist-weak}
Fix any market $\left(I,C,(q_c)_{c\in C},\succsim,\tau\right)$. Let $\succ_c^\tau$ denote the strict priority of $c$ obtained by applying the tie-breaking rule $\tau$ to $c$'s original weak priority $\succsim_c$. Also, let $\succ_c^{*,\tau}(P)$ denote the strict priority of $c$ obtained by applying the tie-breaking rule $\tau$ to $c$'s modified weak priority $\succsim_c^*(P)$. Now, consider when $i\succ_c^\tau j$ and $r_{ic}(P_i)\le r_{jc}(P_j)$. If $i\succ_c j$, the weak rank monotonicity gives $i\succ_c^* j$ and thus $i\succ_c^{*,\tau}(P)\,j$. If instead $i\sim_c j$ and $i\,\tau_c\,j$, the weak rank monotonicity gives $i\succsim_c^* j$ and thus $i\succ_c^{*,\tau}(P)\,j$. Now, go back to our original formulation and treat $\succ_c^\tau$ as the original priority and $\succ_c^{*,\tau}(P)$ as the modified priority. Then, rank monotonicity is satisfied by the above argument. Also, by definition, RD stability and strategy-proofness are satisfied. Then, \Cref{thm:noexist} completes the proof. \qed

\bibliographystyle{apalike} 
\bibliography{citation_v2}
\end{document}